\documentclass[12pt]{iopart}

\usepackage{graphicx}

\usepackage{cite}

\begin{document}

%\preprint{BNL-62600}

\author{D.~E.~Kahana and S.~H.~Kahana}
\address{Physics Department, Brookhaven National Laboratory\\
   Upton, NY 11973, USA}
\eads{\mailto{kahana@bnl.gov}, \mailto{dek@bnl.gov}}
\title{J$/\psi$ Production by Charm Quark Coalescence}

\date{\today}  

\begin{abstract}

Production of $c\bar c$ pairs in elementary hadron-hadron collisions is
introduced in a simulation of relativistic heavy ion collisions. Coalescence
of charmed quarks and antiquarks into various charmonium states is performed
and the results are compared to PHENIX J$/\psi$ Au+Au data. The $\chi$ and
$\psi$' bound states must be included as well as the ground state J$/\psi$,
given the appreciable feeding from the excited states down to the J$/\psi$ via
gamma decays. Charmonium coalescence is found to take place at relatively late
times: generally after $c$($\bar c$)-medium interactions have ceased.  Direct
production of charmonia through hadron-hadron interactions, {\it ie.}  without
explicit presence of charm quarks, occurring only at early times, is
suppressed by collisions with comoving particles and accounts for some $\sim
5\%$ of the total J$/\psi$ production.  Coalescence is especially sensitive to
the level of open charm production, scaling naively as $n_{c\bar c}^2$. The
J$/\psi$ transverse momentum distribution is dependent on the charm quark
transverse momentum distribution and early charm quark-medium interaction,
thus providing a glimpse of the initial collision history.

\end{abstract}

\submitto{\JPG}

\maketitle 

\section{Introduction}

A consistent and successful theoretical description of Au+Au and D+Au,
collisions at RHIC ($s^{1/2}=200$A GeV), including pseudorapidity and
$p_\perp$ spectra, as well as transverse momentum suppression and elliptical
flow, has been developed with the simulation
LUCIFER~\cite{luciferJPG1,luciferJPG2}. LUCIFER models ion-ion collisions by
the early production -- immediately after the completion of an initial phase
of high energy interactions -- and subsequent interaction of a fluid of
pre-hadrons, having properties not unlike those of the $q\bar q$ dipoles used
by many authors~\cite{boris1}. In the present work, we introduce charmed
quarks explicitly into the picture, produced via hadronic collisions in the
initial phase. Coalescence of charm quarks and anti-quarks into charmonia, which
can also be viewed as heavy pre-hadrons, is calculated in a model we first
used for deuteron production in Au+Au collisions at the
AGS~\cite{AGSdeuterons}, adapted here to treat $c\bar c \rightarrow$
charmonium.  Only a small percentage of charm quarks are found to coalesce
into bound charmonia {\it ie.}  J/$\psi,\chi$ and $\psi$'; the remainder of
such quarks appear ultimately as open charm mesons.

We find that the situation with respect to charmonium production present at
the SPS is very much altered by the higher energy available at
RHIC. Coalescence of $c,\bar c$ pairs into charmonium increases in importance,
occurs later and coalesced charmonia are not as subject to comover
suppression. At SPS, comover suppression was sufficient to explain the
observed suppression of J$/\psi$ yields relative to the expected direct
production.  At RHIC energies comover suppression of directly produced
charmonia is found to be even larger, due to the increased particle numbers
and densities. This will be spelled out in more detail in what follows.

In elementary hadronic interactions open charm and charmonium production are
constrained by existing data at a variety of
energies~\cite{PhysRevD.66.092001,1742-6596-50-1-047,1997PhLB..410..337N,
  PhysRevD.55.3927,PhysRevLett.60.2121} including the data taken at RHIC
~\cite{PhysRevLett.96.032001,PhysRevLett.97.252002,PhysRevLett.94.062301,Zhang:2005hi}.
Such cross-section information is used as a basic input for the LUCIFER
simulation of A+A collisions and of course does not invoke adjustable
parameters; there is little dependence on the explicit functional forms chosen
to describe the input data.  Indeed, for the purposes of studying charmonium
production the detailed simulation can be viewed as providing only a
background for the coalescence calculation, which to a large extent then
stands on its own.  The generation of free quarks and the wave functions for
the charmonium states into which they coalesce are taken, respectively, from
measured open charm production cross-sections and from phenomenological
analyses of charmonium electromagnetic decays.

\section{Simulation Dynamics}

The simulation dynamics has already been extensively discussed in earlier
publications, most recently in References ~\cite{luciferJPG1,luciferJPG2}, and
will only be briefly outlined here. LUCIFER is a two stage
simulation~\cite{PhysRevC.58.3574,PhysRevC.63.031901}. In stage I the incoming
target and projectile nucleon interactions are tracked, while in stage II the
produced pre-hadrons interact and decay in a standard relativistic cascade
model. The time history of all the collisions recorded in stage I sets up the
geometry and initial conditions for stage II.  Basic inputs for the simulation
are measured hadron-hadron
cross-sections~\cite{Alner:1986is,Albajar1990261,PhysRevD.41.2330}, and
rapidity, transverse momentum and particle multiplicity distributions, with
the last taken to conform to KNO~\cite{Koba1972317} scaling.  Wherever
hadronic cross-sections are unknown we employ quark counting to estimate
their magnitudes, implying, for example, that the pre-meson/pre-meson
cross-section would be $\sim 4/9$ the nucleon/nucleon cross-section at the
equivalent center of momentum energy, {\it ie.} $\sim 20-25$ mb numerically.

Pre-hadrons, being in principle off shell, do not correspond directly to
hadron resonances listed in the particle data book. We choose their masses
from a Gaussian distribution, centered at $m_0$ with width $m_0/4$, where
$m_0$ is dependent on the particular type of pre-hadron, that is to say,
$\rho$-like (non-strange), or $K^*$-like (strange). For non-strange (strange)
pre-hadrons we take $m_0\sim 700 (950)$ MeV. Small changes in these masses do
not alter results since one must always readjust mulitplicities and momentum
distributions to fit known two body data.

All produced pre-hadrons are placed, at the end of stage I, uniformly inside
the overlap region of the colliding nuclei. The pre-hadrons are then allowed
to evolve longitudinally with the already assigned momenta, for a fixed time,
numerically on the order of $0.30$ fm/c. The total multiplicity of pre-hadrons
is limited so that, given normal mesonic sizes $\sim 0.55$ fm/c appropriate to
meson-meson cross-sections, pre-hadrons do not overlap physically. The implied
limitation in density is consonant with pre-hadrons existing as distinct
objects only after becoming separated
spatially~\cite{PhysRevLett.32.957,1861519}. One may conclude from this that
the pre-hadronic matter is, when first created, something like an
incompressible fluid, as in earlier calculations with
LUCIFER~\cite{PhysRevC.58.3574}.

Stage II is a straightforward two body cascade with collisions among the
pre-hadrons taking place at considerably lower energy than stage I.
Pre-hadrons collide, resulting in the production of more pre-hadrons and
changes in momentum distributions. In addition pre-hadrons are allowed to
decay by the emission of pions, the decay chains ending at stable mesons and
baryons. The pre-hadron decay time at rest $\sim$ 1.0 fm/c., could also be
viewed as a hadronisation time.  Final state interactions in stage II are the
principal agent suppressing the production of high $p_\perp$ particles (jet
suppression)~\cite{luciferJPG1,Kahana:2004pd} in Au+Au and D+Au collisions.

Central to the dynamics are the two time scales $t_p$, $t_f$ defined in the
rest frame of the two colliding nuclei.  These times are the production and
hadronisation times for pre-mesons.  This has been described in detail in
previous works ~\cite{luciferJPG1,luciferJPG2}. For clarity we repeat and
condense those arguments here. Colour neutral pre-hadrons are produced
perturbatively at time $t_p$, by a process that can be viewed as the
coalescence of a target quark struck by a gluon with an anti-quark that was
generated in a slightly later an separate pair creation event.  Integrating
over soft gluon radiation from the initial quark, with an accompanying hard
scale $Q$ yields~\cite{boris1} a pQCD estimate for the production time:

\begin{equation}
t_p \sim
\frac{E_q}{Q^2} (1-z_h),
\end{equation}
where $z_h=E_h/E_q$ is the fraction of energy imparted to the hadron, and
$E_q$ and $E_h$ are the quark and pre-hadron energies. The hadronisation time
$t_f$, which is approximated as
\begin{equation}
t_f \sim \frac{E_h}{\Lambda^2_{QCD}},
\end{equation}
\noindent
is far longer than the production time $t_f$, given that the
$\Lambda_{\textrm{QCD}}$ is about 200 GeV.

In the center of mass frame these times translate to $\tau_{p,f}\sim
\gamma^{-1} t_{p,f} $. This implies that the pre-hadrons are produced early on
and interact for an extended period before decaying. These times, which are
the main free parameters in the simulation, have already been determined in
earlier work~\cite{luciferJPG1} and were not altered for the purposes of the
present calculation.

We have previously noted the suggestion of Molnar and
Voloshin~\cite{PhysRevLett.91.092301, Molnar:2000yv, Molnar:2001nk} that
coalescence of $q\bar q$ pairs could help to explain elliptical flow in
ion-ion collisions.  Pre-hadrons, in our simulation, have interaction
cross-sections sufficiently large to explain the surprisingly large observed
flow~\cite{luciferJPG2}: flow being the only truly collective variable
observed at RHIC. The production mechanism of pre-hadrons, in view of their
$q\bar q$ pair structure, is also akin to coalescence.

\section{Coalescence of Charmed Quarks.}

Due to the strong energy dependence of open charm production cross-sections,
charmed quarks are mainly produced early, in the high energy collisions
occurring in stage I of the simulation, and to a much lesser extent, early in
stage II. In our treatment charmed quarks then propagate and interact with
pre-hadrons in stage II. Stage II of the simulation then runs to its natural
conclusion, which occurs when no more two-body collisions with CM energy above
a minimum cutoff are detected.

Coalescence of charmed quarks and antiquarks into charmonium states is then
accomplished by an afterburner algorithm which searches through the final set
of particles for $c\bar c$ pairs and joins them, or not, on the basis of an
estimate of the quantum mechanical overlap probability of the $c\bar c$ pair
with the charmonium final state. Pairs which coalesce are therefore
necessarily closely correlated in relative position and momentum.  This
approach to coalescence was used successfully, as stated above, to calculate
deuteron yields in much lower energy Au+Au collisions at the
AGS~\cite{AGSdeuterons}.  We find at RHIC, as was the case for J/$\psi$
production at CERN
SPS~\cite{Kahana:1999qp,PhysRevC.59.1651,1742-6596-50-1-047,Abreu:2000ni,Bordalo2002127,Armesto2002583},
that the early, direct charmonium production is strongly suppressed by the
interaction with baryons in stage I, as well as collisions with co-moving
pre-mesons in stage II.  In our picture, most observable charmonia at RHIC
arise from coalescence of $c\bar c$ pairs at late times during the ion-ion
collision.

The coalescence ``afterburner'' tests each $c\bar c$ pair for possible merging
using recorded information about the history of the position and momentum of
the $c$ and the $\bar c$ to estimate wavepacket sizes for each. For every
$c\bar c$ pair considered, the distance of closest approach is determined,
assuming the closest approach time occurs in the future. At the time of
closest approach for most pairs interactions with the medium have virtually
ceased, due to a rapid drop in the density of comoving particles. We call the
procedure of following the pair in their passage through the interacting
medium and the use of that history to determine initial $c\bar c$ wave packets
dynamic coalescence. The overlap integral is then constructed between the $c$
and $\bar c$ wavefunctions and an appropriately defined bound state charmonium
wavefunction.  The overlap generates a probability for coalescence which is
used in the calculation, by Monte-Carlo, of the charmonium yield.

It's also possible to calculate separately direct production of charmonium in
our model, using the known elementary hadron-hadron production cross-sections,
where these are available, or being guided by quark counting, where they are
not available. At RHIC we find that the ensuing breakup of charmonia, if
produced directly and early during ion-ion collisions, is very significant,
even greater than at SPS.  The mechanism of direct production followed by
co-mover suppression successfully
described~\cite{Kahana:1999qp,PhysRevC.59.1651} the J$/\psi$ suppression as a
function of $E_\perp$, observed at the
SPS~\cite{1742-6596-50-1-047,Abreu:2000ni,Bordalo2002127,Armesto2002583}.
Interestingly it was also necessary, at SPS, to include all bound charmonia in
the theory.

At RHIC we find that the number of bound charmonia formed by coalescence rises
rapidly with the initially produced number of charmed quarks. Charmonium
production scales with $\sim n_{c\bar c}^{2.5}$ where $n_{c\bar c}$ is the
number of charm quarks produced per ion-ion collision: but the suppression of
direct charmonium production at the same time increases relative to the SPS.
In fact in the RHIC environment we find that the direct production mechanism
results in considerably less (about a factor of $20$) charmonium than
coalescence. Since the two processes are to a large extent independent,
sufficient accuracy can be obtained by simply adding the two contributions.

\section{Coalescence: Details of Calculation}

Coalescence of $c\bar c$ into charmonia is achieved by calculating the overlap
integral between wavepackets composed of plane waves for the incoming $c$ and
$\bar c$, with a wavefunction for the outgoing charmonium bound state, which
is the product of a wavepacket consisting of plane waves in the center of
momentum coordinates, and a bound state wavefunction in the relative
coordinates. The assumption is made that the center of mass $3$-momentum is
conserved. Explicitly, we write for the wavefunctions of the $c$ and $\bar c$:

\begin{equation}
\psi_{c,\bar c} ({\bf x}_{c,\bar c}) = \left({\frac{1}{\pi \sigma^2}}\right)^{3/4} \exp \left( 
- \frac{({\bf x}_{c,\bar c}-{\bar {\bf x}}_{c,\bar c})^2}{2\sigma^2_{c,\bar c}}\right)
\exp(i{\bar {\bf p}}_{c,\bar c}\cdot {\bf x}_{c,\bar c}).
\end{equation}

\noindent And for the wavefunction of the charmonium we write:

\begin{equation}
\Psi_{c\bar c} ({\bf x}_c,{\bf x}_{\bar c}) = \Phi_{{\bar {\bf
      P}},{\bar {\bf R}}} (\bf R) \phi_{c\bar c}(\bf r),
\end{equation}
\medskip
\noindent where

\begin{equation}
\Phi_{ {\bar {\bf P}}, {\bar {\bf R}}}(\bf R) = \left({\frac{1}{\pi \Sigma^2}}\right)^{3/4} 
\exp\left(-\frac{({\bf R}-{\bar {\bf R}})^2}{2\Sigma^2}\right) \exp (i{\bar
  {\bf P}} \cdot {\bf R}),
\end{equation}

For the sake of simplicity, we employ three-dimensional oscillator states for
the relative wavefunctions, with radius parameters selected to agree with the
rms radii extracted from Eichten {\it et al.}
~\cite{PhysRevD.17.3090,PhysRevD.21.203}.  The Cornell wave functions for the
0s, 0p and 1s states are to a fairly good approximation harmonic oscillator
states, originating, as they do, from a coulomb+linear model potential:

\begin{equation}
V(r) = -\frac{\kappa}{r} + \frac{r}{a^2},
\end{equation}

\noindent with $\kappa$ dimensionless and $a$ possessing the dimensions of
$[L]$.  The strength of the Coulomb coupling $\kappa$ is related to the strong
coupling at short distances, while the strength of the linear coupling $a$, is
proportional to the Gaussian radius parameter and the linear potential, of
course, is confining.  So, for the (0s) J/$\psi$ we use

\begin{equation}
\phi^{0s}_{c\bar c} (r) = (\pi \alpha^2)^{-3/4} \exp(- r^2/2\alpha^2),
\end{equation}

\noindent as the relative wavefunction, with $\alpha$ chosen appropriately.

The approach to charmonium wavefunctions based on the Cornell potential is of
course non-relativistic, but may still be reasonably accurate, given the large
charm quark mass $\sim 1.5-1.6$ GeV.  The same charm quark mass is used in the
LUCIFER simulation. For the incoming wave packets the total and relative
momenta are related by:

\begin{equation}
{\bf {\bar P}}=({\bf {\bar p}}_c + {\bf {\bar p}}_{\bar c}),
\end{equation}

\begin{equation}
{\bf \bar p}=\frac{({\bf {\bar p}}_c - {\bf {\bar p}}_{\bar c})}{2},
\end{equation}
\noindent while the relative separation between c and $\bar c$
wavepacket centers is
\begin{equation}
{\bf \bar r}=({\bf {\bar x}}_c - {\bf {\bar x}}_{\bar c}).
\end{equation}
\noindent and for completeness, the total (CM) position is given
by:
\begin{equation}
{\bf \bar R}=\frac{({\bf {\bar x}}_c + {\bf {\bar x}}_{\bar c})}{2}.
\end{equation}
 
\noindent The final CM wave function is taken identical to its incoming form,
          {\it ie.} total $3$-momentum is preserved and we take $2\Sigma^2 =
          \sigma_c^2 = \sigma^2_{\bar c}$. The overlap integral

\begin{equation}
P = \left\vert \langle \psi_c \psi_{\bar c} \vert \Phi_{\bar P, \bar R}\phi_{c\bar c}\rangle \right\vert^2,
\end{equation}

\noindent is constructed. After evaluation, the explicit probability for
formation of 0s-state, {\it ie.} the J$/\psi$~\cite{AGSdeuterons} is found to
be:

\begin{equation} 
P^{0s}=A\exp(-B),
\end{equation}
\noindent where

\begin{equation}
A=\left[\frac{4\alpha\sigma}{\sqrt{2}(\alpha^2+2\sigma^2)}\right]^3 \textrm{and}\,\,\,\,
B=\left[\left(\frac{2\alpha^2}{\alpha^2+2\sigma^2}\right)\,\,{\bf \bar
    p}^2+\left(\frac{1}{\alpha^2+2\sigma^2}\right)\,\,{\bf \bar r}^2\right].\end{equation}

\noindent Here $\alpha$ is the size parameter for the J$/\psi$(0s) and
$\sigma$ is that for the incoming free charm wave packets.

Coalescence probabilities for the $\chi$(0p) and $\psi'$(1s) charmonium
states can be generated from $P^{0s}$ by taking appropriate derivatives with
respect to momentum and position, since we are using harmonic oscillator
wavefunctions as approximations to the Cornell group
wavefunctions~\cite{PhysRevD.17.3090,PhysRevD.21.203}, and thus have,
ultimately, Gaussian integrals to perform. After some algebra, one obtains for
the p-state:

\begin{equation}
P^{0p} = \left(\frac{2}{3}\right)\,P^{0s}\,\left[{\bf \bar p}^2\sigma^2+\frac{{\bf \bar r}^2}{4\sigma^2}\right]
\left(\frac{2\alpha\sigma}{\alpha^2+2\sigma^2}\right)^2 ,
\end{equation}

\noindent and for the 1s-state (with one node):

\begin{equation}
\fl P^{1s}=
\left(\frac{2}{3}\right)
\,
P^{0s}
\,
\left[
\left\lbrace
2\left[
{\bf \bar p}^2(\sigma)^2-\frac{{\bf \bar r}^2}{4(\sigma)^2}\right]
\left(\frac{\alpha\sigma}{(\alpha^2+2\sigma)}\right)^2
+ 3\left(\frac{\alpha^2-2\sigma^2}{\alpha^2+2\sigma^2}\right)
\right\rbrace^2
+\left(\frac{2\alpha\sigma}{\alpha^2+2\sigma^2}\right)^4
({\bf \bar p}\cdot{\bf \bar r})^2\right].
\end{equation}

The Cornell group~\cite{PhysRevD.21.203} determined properties of the
J/$\psi$, $\chi$ and $\psi'$ from the decays of these states, electromagnetic
and otherwise. In particular, the rms radii were found to be $0.47$ fm, $0.74$
fm and $0.96$ fm respectively, the latter two being somewhat larger than
expected for oscillator wavefunctions with a common radius parameter $r_0$.
To better approximate the Cornell wavefunctions using simple oscillator
wavefunctions, we chose different radius parameters for the different states:
$r_0(\psi)=0.384$ fm, $r_0(\chi)=0.49$ fm and $r_0(\psi')=0.51$ fm.

The enhancement of the spatial size of the higher states over pure
oscillators, suggests that coalescence will populate these states at a higher
rate, and this is indeed found to be the case.  However, the increased
relative momentum dependence is somewhat quenched, since the average $p_\perp$
in Au+Au collisions is generally larger than the bound state relative momentum
content $|{\bf \bar p}| \leq 1/r_0 \sim 500$ MeV.  We expect to find
successful coalescence in charm pairs for a distribution decidedly cutoff at
higher $|{\bf \bar p}|$ and hence higher $p_\perp$. The $\chi$ and $\psi$
states feed down to the J$/\psi$ by their gamma decays, so that the yield of
J$/\psi$ due to coalescence is enhanced by the use of larger oscillator radii
for these states. The overall production of J$/\psi$ is more than doubled by
feeding from the higher states.

\section{Results}

The measured $NN\rightarrow c\bar c$ cross-section over a range of energies is
a necessary input to the coalescence calculation.  We fitted existing data at
several energies, using in particular the
PHENIX~\cite{PhysRevLett.96.032001,PhysRevLett.97.252002} measurements at 200
GeV.  STAR~\cite{PhysRevLett.94.062301,Zhang:2005hi} and PHENIX results for
$\sigma_{c\bar c}^{pp}$ differ significantly, with little room for
convergence, even within the rather large experimental errors quoted. Our
calculation of the charmonium yields uses the most recent PHENIX $pp$
cross-section $\sigma^{pp}_{c\bar c}=0.57 \pm 0.21$
mb~\cite{PhysRevLett.97.252002}.  STAR reports~\cite{PhysRevLett.94.062301} a
somewhat higher cross-section $\sigma^{pp}_{c\bar c}$, for minimum bias,
inferred from the D+Au measurement of $\sigma_{c\bar c}^{NN}= 1.40\pm 0.22$
mb, while extracting $\sigma_{c\bar c}^{NN} = 1.11\pm 0.43$ mb, from the
Au+Au~\cite{Zhang:2005hi} measurement. The STAR value obtained from Au+Au is,
however, sensitive to theoretical input, so that this result should probably
not be used to minimise the discrepancy between the collaborations.

For consistency we compare our calculations, for the most part, to PHENIX
data, and use their estimated cross-section for elementary charm production.
On occasion though, we make use of the STAR results: their $p_\perp$
dependence for open charm importantly confirms the conclusion drawn from the
PHENIX J$/\psi$ data, {\it ie.}  that charm production and interaction with
the dense medium in heavy ion collisions takes place early on.

Future experiments hopefully will settle the apparent
discrepancy~\cite{dunlop} between PHENIX and STAR measurements of open charm
production, which is likely the largest potential source of error in our
calculation of the charmonium yield due to $c \bar c$ coalescence, considering
the great sensitivity of coalescence to the number of $c\bar c$ pairs.
Naively coalescence is expected to vary with the square of $n_{c\bar c}$; in
fact an even more rapid dependence results from the considerably less than
3-dimensional emission of plasma in A+A collisions, since in general $p_\perp
\ll p_l$.

Employing the formation probabilities shown in the last section and applying
the measured branching ratios for the decay into J/$\psi$ of the higher
charmonium states, we obtain in a straightforward manner the overall yield of
J$/\psi$.  These results are presented in a series of figures, beginning with
global rapidity and transverse momentum distributions and their dependences on
centrality.  Figure~(1) compares rapidity distributions from the simulation,
for both J$/\psi$ and open charm to those from
PHENIX~\cite{PhysRevLett.98.232301}. Figure~(2) is a similar comparison for
$p_\perp$ dependent invariant cross-sections.  Both figures are for central
collisions of Au+Au.

Figure~(3) displays the centrality dependence of the mid-rapidity yield of
J$/\psi$ in Au+Au collisions, again comparing LUCIFER results to PHENIX
measurements~\cite{PhysRevLett.98.232301}. The simulated yields parallel the
experimental results well and of course these variations arise directly
from the underlying geometry of ion-ion collisions.

The STAR collaboration reconstructed $D^0$ mesons from their mesonic
decays and thus directly measured~\cite{Zhang:2005hi} open charm production,
rather than using `non-photonic leptons' as a proxy for open charm. They
measured the variation for low $p_\perp$  open charm.  Figure~(4)
compares STAR results to the LUCIFER calculations using an appropriate
branching ratio $r = 0.52$ for $D^0$ decay.

We alluded above to the significant dependence of the overall coalescence
probability for J$/\psi$ on the distributions of relative momentum and
$p_\perp$ in the theoretical simulation.  This is nicely illustrated  in
Figure~(5) where the $|{\bf \bar p}|$ distributions for all $c\bar c$ pairs
and separately for coalesced pairs are shown. Clearly, $c\bar c$ pairs with $|{\bf \bar
  p}|\ge 500$ MeV/c are hindered from merging.

The rapidity distribution of directly produced J$/\psi$'s, as calculated in
LUCIFER, is shown in Figure (6). This calculation proceeds simply given
known cross-sections for charmonium production in $NN$ collisions. The various
charmonium states are directly and promptly created -- for the most part in
initial $NN$ collisions in stage I of the cascade due to the high energies
available and the rapid energy loss -- but charmonia can also be created to
some extent by the most energetic pre-hadron collisions that take place in stage
II.  These directly produced charmonia are then allowed to interact and
potentially be broken up in both stages of the simulation. This approach was
employed and extensively described in our earlier work on J$/\psi$ suppression
~\cite{Kahana:1999qp} at the SPS, where the production of open  (as well
as hidden) charm was appreciably smaller and $c\bar c$ coalescence  far
less important.

Direct production turns out to be a minor contribution at RHIC energy:  
direct yield is only some $5\%$ of 
coalescent yield, for  central collisions. This is because of the very
efficient breakup in collisions with comovers, of any promptly produced
charmonia given the higher particle densities existing at RHIC energy relative
to the SPS.  The rapidity distribution of directly produced
charmonia differs very little from that of  coalescence and
we show the summed rapidity distributions in Figure~(1). It seems that the
present simulation describes the known RHIC data adequately, both for open
charm and for J$/\psi$, with coalescence as the major mechanism of charmonium
production.

\subsection{Charm Quark Interactions}

Introduction of charm quarks into the simulation opens a theoretical window on
the early collision environment.  The interaction of free charm quarks with
the dense medium, initially with nucleons in stage I, thereafter with
pre-hadrons in stage II of the simulation, is another input into the
coalescence calculation. Once again we employed quark-counting to obtain an
estimate of pre-hadron/quark interaction cross-sections, as a starting
point. That is, we took the interaction cross-section for charm quarks ($c$)
with pre-hadrons ($h$) to be  $\sigma(c h) = \frac{1}{2}
\sigma(\pi h)$.

One might expect, considering the lore on colour transparency, that a free
quark interacting with a colourless pre-hadron would be subject to greater
screening, suggesting that the quark-counting estimate invoked for
hadron-hadron interactions might over-estimate the magnitude of the
interaction, with the heaviness of the charm quark being a further variable.
Perturbative QCD might offer some guidance, but for this exploratory
calculation we explored the effects of differing $c$($\bar c$)-hadron
interaction strengths characterized by a dimensionless parameter s ($\sigma(c
h) = \frac{1}{2} \, s \, \sigma(\pi h)$).  Specifically we show results for $s
\in [0,1]$, where the limit s=1 corresponds to the quark counting estimate of
the cross-section, and s=0 corresponds to free-streaming charmed quarks.
Figure~(1) displays one such comparison.  Within the errors permitted by direct
measurements of J/$\psi$.~\cite{PhysRevLett.98.232301} and certainly within
the much larger range of uncertainty allowed by PHENIX and
STAR~\cite{PhysRevLett.97.252002,PhysRevLett.94.062301} measurements of
$\sigma^{pp}_{c \bar c}$, pure quark counting $s=1$, appears to give a more
than adequate description, perhaps validating the overall approach.

The coalescence yield in Figure~(7) shows an interesting transverse momentum
variation with the charm quark/medium interaction strength $s$. If the charm
quarks are allowed to free stream, then the J/$\psi$ $p_\perp$ distribution
falls too quickly with $p_\perp$, while for $s\neq 0$ the interaction adds
significant transverse momentum to the c-quarks, thus increasing the $p_\perp$
of the coalesced J$/\psi$ and leading to reasonable agreement with the measure
$p_\perp$. Open charm in A+A collisions at RHIC apparently is mostly produced
early on in the ion collisions, during the initial full energy nucleon-nucleon
collisions of stage I. To a lesser extent there is also production of open
charm from the few higher energy pre-hadron/pre-hadron collisions that occur
early in stage II. So the actual $p_\perp$ variation in the
PHENIX~\cite{PhysRevLett.98.232301} charmonium data provides information
on the initial state of the heavy ion collisions.

\section{Conclusions and Discussion}

All in all the simulation represents known RHIC data well, and the results
suggest that coalescence of $c\bar c$ pairs is the dominant mechanism for
charmonium production in heavy ion collisions at RHIC.  The strong sensitivity
of coalescence yields to the elementary production of free charm is evident. A
reasonable picture emerges for rapidity and transverse momentum distributions
of J$/\psi$ as well as for the variation of the yield with
centrality. Throughout, we emphasized PHENIX data in our comparisons;
underpinning our calculation of the magnitudes of the J/$\psi$ yield by the
PHENIX open charm production cross-section $\sigma^{pp}_{c\bar c}$ at 200A Gev
though, of course, charm production cross-sections at lower energy were also
require, and used. The coalescence calculation clearly could be relatively
easily extended to the treatment of bottomonia.

Coalescence must be taken seriously as a mechanism for production of heavy
quarkonia in higher energy heavy ion collisions; direct production of
charmonium, though  successful  at the SPS, was not
found  to be a viable explanation for J/$\psi$ yields at RHIC. The nature
of the bound state wave functions for the coalesced particles plays an
important role, as does the explicit inclusion of all of the charmonium bound
states. These factors greatly enhance the likelihood of J$/\psi$ formation.
Knowledge of the actual structure of bound $c\bar c$ states as presented by
Eichten {\it et al.}~\cite{PhysRevD.17.3090,PhysRevD.21.203} and supported by
the plethora of data on charmonium decays, provides a solid foundation for the
conclusions drawn, for both the absolute coalescence yields and the momentum
distributions.  An improved determination of elementary open charm production
could perhaps turn the heavy ion J$/\psi$ measurements into a spectroscopic
tool.

Further, as stated above, the $p_\perp$ dependence of the J$/\psi$ yield from
coalescence is unambiguously tied to the magnitude of early charm quark
production and to the strength of the charm quark interaction in the dense
medium: and thus charmonium date provide, perhaps, the lone hadronic signal of
the initial collision environment. More generally, heavy quarkonia may provide
alternative signals of interaction within the dense medium present at early
stages of ion-ion collisions. The approach pursued here is consonant with our
overall model of the `pre-hadron' dynamics~\cite{luciferJPG1,luciferJPG2}
employed in LUCIFER to extract single particle spectra in general, that is to
say: we could equally well consider charmonia as heavy pre-mesons.  By
explicitly including charm quarks as participants in the cascade we obviously
recognize that at some level pQCD, which surely underlies all hadronic
dynamics, must be considered. Our previous
work~\cite{luciferJPG1,luciferJPG2,PhysRevC.59.1651,Kahana:1999qp} gave
credence to the notion of pre-mesons: these being simply correlated $q\bar q$
pairs arising very likely by coalescence of their quark and anti-quark
components~\cite{Molnar:2000yv,Molnar:2001nk}.  In any case coalescence for $c
\bar c$ constitutes one estimate of a soft QCD process, {\emph ie.}  the
fragmentation of charmed quarks.

At RHIC energies and above, and for more central ion-ion collisions,
coalescence of $c\bar c$ dominates the production of charmonia. Heavy
quarkonia may well be key to understanding the early collision history when
ions are collided at the LHC.  But the large number of free quarks and gluons
expected to be produced at such energies, on the basis of jet statistics in
$p\bar p$ collisions, means that partons are more likely to be directly
involved. For example the large gluon numbers created in ion collisions may
well enhance the likelihood of thermalisation.

\ack
This  manuscript  has  been  authored  under  the  US  DOE  grant
NO. DE-AC02-98CH10886. One of  the authors (SHK) is also grateful
to  the  Alexander von  Humboldt  Foundation,  Bonn, Germany  for
continued support and H.-J.  Pirner, University of Heidelberg for
hospitality. Additional thanks are due Johanna Stachel for  a
highly relevant  conference presentation (HEP-2006,  Valparaiso,
Chile).

\section*{References}
\bibliography{jpsi_coal}
\bibliographystyle{iopart-num}

\begin{figure}
%\vbox{\hbox to\hsize{\hfil
%\epsfxsize=6.1truein\epsffile[0 0 561 751]{figs/PUB.Journal.dndy.ps}
%\hfil}}

\includegraphics*[trim= -45 0 -10 -10,scale=0.95]{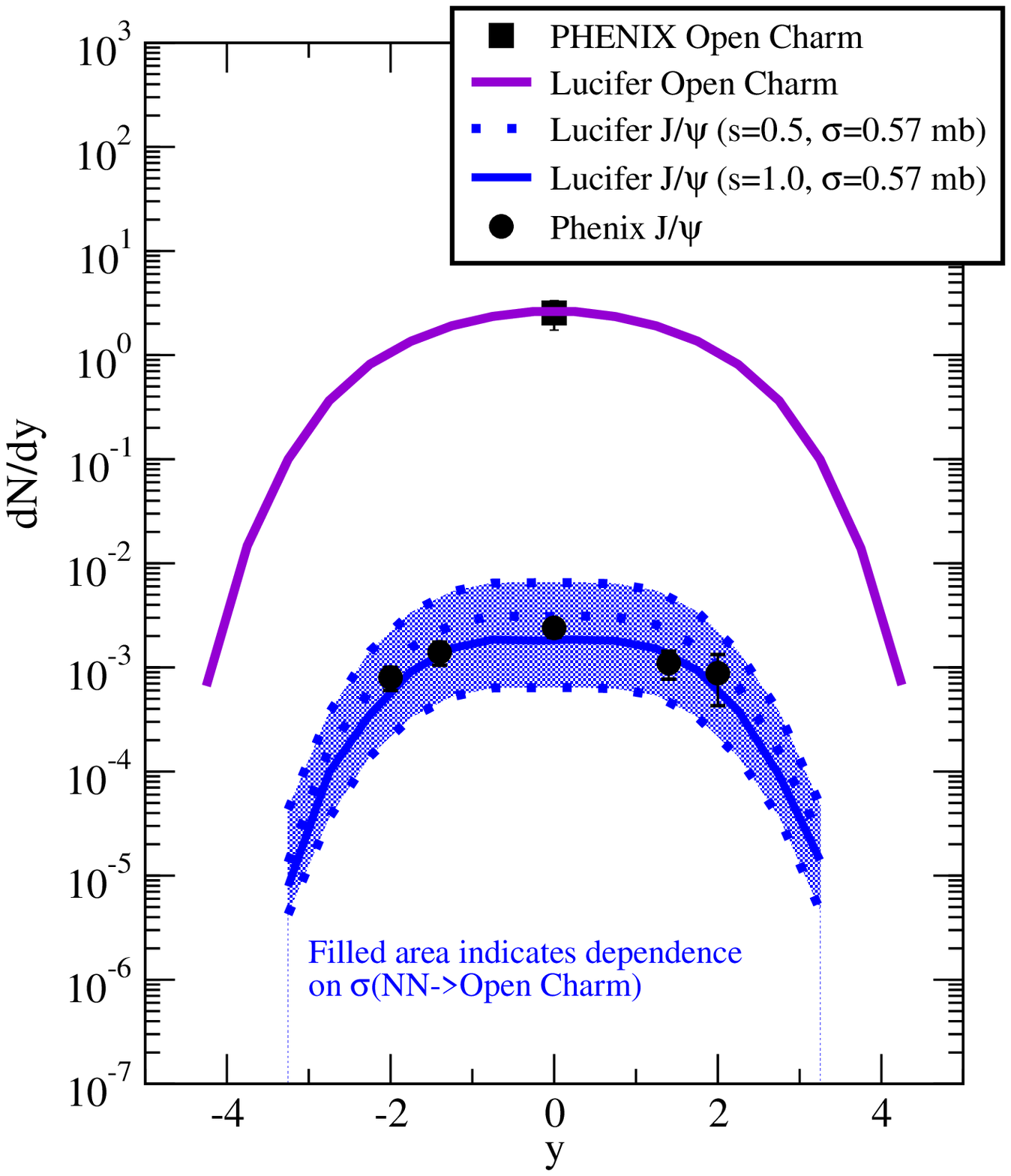}
\caption[]{Comparison of Au+Au $J/\psi$ central production ($0-20\%$) and open
  charm rapidity distributions from LUCIFER and PHENIX.  The large theoretical
  variation in overall coalescence yield that results from the stated PHENIX
  errors in open charm production $pp\rightarrow c\bar c$ at 200 GeV, is
  shown.  We chose the most recent PHENIX value ($\sigma^{pp}_{c\bar c} =
  0.57$ mb) for open charm production~\cite{PhysRevLett.97.252002} in
  estimating the yield of charmonium from coalescence. The factor $s$ modulates 
  the  charmed quark-medium interaction, shown here for the range $(0.5,1.0)$
  with $s=1$ corresponding to quark counting (see later text discussion).}
\label{fig:Fig.(1)}

\end{figure}
\clearpage

\begin{figure}
%\vbox{\hbox to\hsize{\hfil
%\epsfxsize=6.1truein\epsffile[0 0 561 751]{figs/PUB.Journal.d2ndydpt.ps}
%\hfil}}
\includegraphics*[trim= -65 0 -10 -10,scale=1.0]{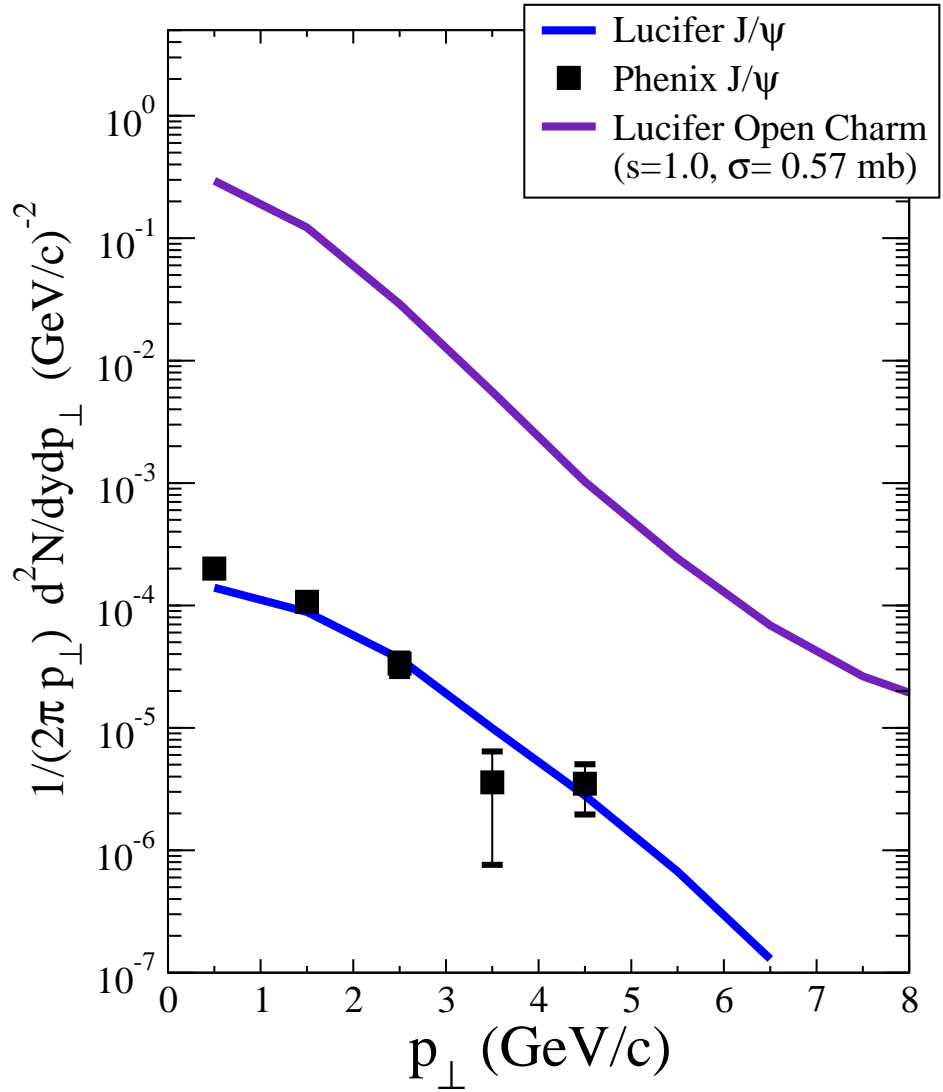}
\caption[]{Comparison of J$/\psi$ production in $20$\% central Au + Au
  collisions: LUCIFER versus PHENIX.  The marked variation in overall
  coalescence normalisation with stated PHENIX errors in the $\sigma^{pp}_{c
    \bar c}$ seen in Figure~(1) is also present here, though it's not
  depicted. The coalescence model gives an adequate description of the
  magnitude and behaviour of the J/$\psi$ transverse momentum
  distribution~\cite{PhysRevLett.97.252002}.}
\label{fig:Fig.(2)}
\end{figure}
\clearpage

\begin{figure}
%\vbox{\hbox to\hsize{\hfil
%\epsfxsize=6.1truein\epsffile[0 0 561 751]{figs/PUB.Journal.Centrality.ps}
%\hfil}} \includegraphics*[trim= -65 0 -10 -10,scale=1.0]{figs/fig3.eps}
\includegraphics*[trim= -65 0 -10 -10,scale=1.0]{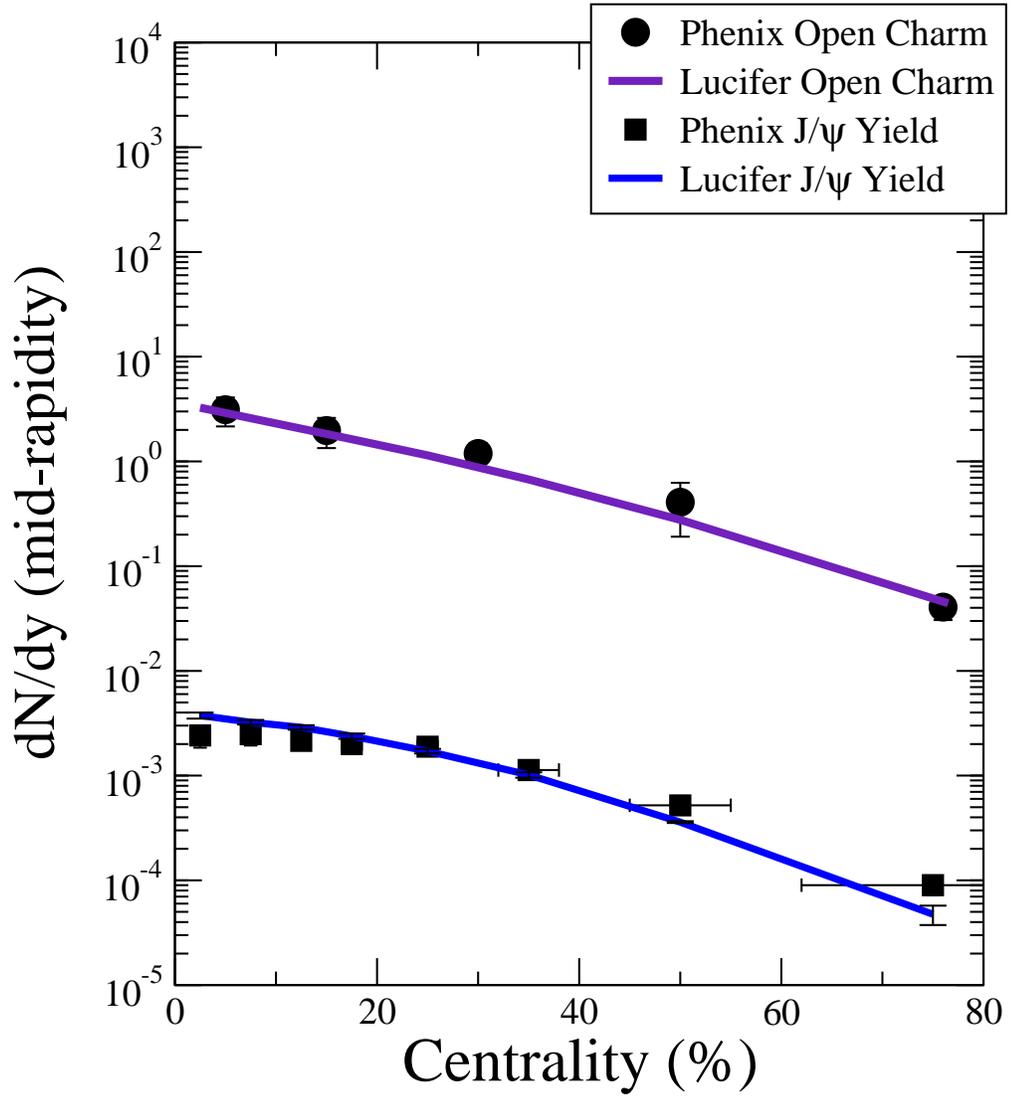}
\caption[]{Comparison of the centrality dependence of J/$\psi$ and open charm
  production: LUCIFER vs PHENIX. A good agreement between the simulation and
  experiment is evident, mostly resulting from the geometry of the ion-ion
  collisions.}
\label{fig:Fig.(3)}
\end{figure}
\clearpage

\begin{figure}
%\vbox{\hbox to\hsize{\hfil
%\epsfxsize=6.1truein\epsffile[0 0 561 751]{figs/PUB.Journal.LUCvsSTAR.charmpt.ps}
%\hfil}} \includegraphics*[trim= -65 0 -10 -10,scale=1.0]{figs/fig4.eps}
\includegraphics*[trim= -50 0 -10 -10,scale=1.0]{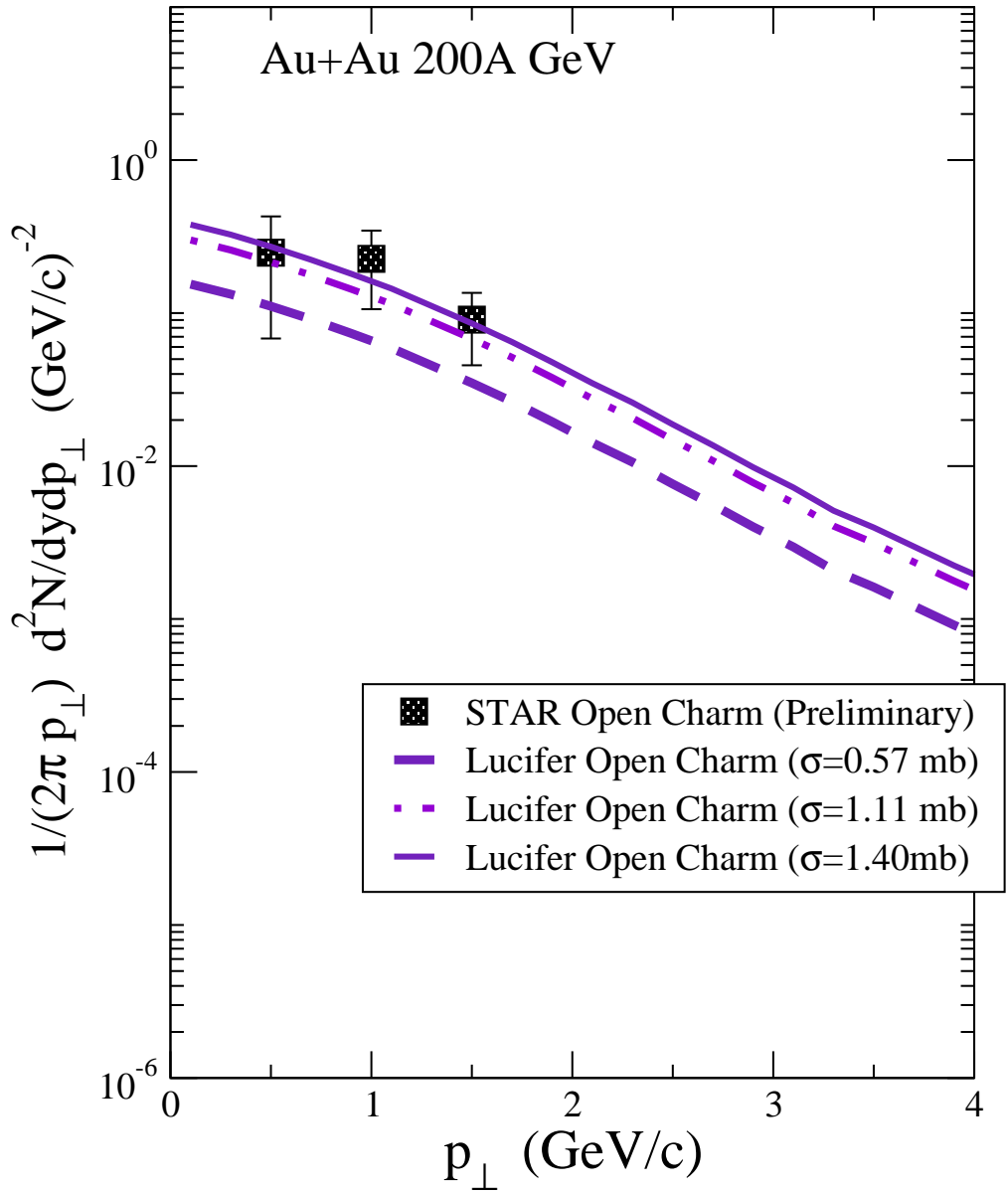}
\caption[]{Open charm invariant cross-section. LUCIFER vs STAR: Analysis of
  $D^0$ production enabled STAR to obtain transverse momentum distributions
  for $c\bar c$ production in Au+Au collisions. Comparison of LUCIFER yields
  are made to this data for various values of $\sigma^{NN}_{c\bar c}$
  including the estimates from STAR D+Au (1.4mb) and PHENIX(0.57mb), and the
  value extracted by STAR from Au+Au collisions (1.1mb). Clearly, the
  calculated differences for open charm production in Au+Au are not as strong
  as the theoretical variation seen in coalesced J/$\psi$. Interestingly the
  STAR comparison to theory is self-consistent, given their determination of
  the $NN\rightarrow c\bar c$ and Au+Au cross-sections.}
\label{fig:Fig.(4)}
\end{figure}
\clearpage

\begin{figure}
%\vbox{\hbox to\hsize{\hfil
%\epsfxsize=6.1truein\epsffile[0 0 561 751]{figs/PUB.Journal.relativemomentum.ps}
%\hfil}}
\includegraphics*[trim= 0 0 -10 -10,scale=1.0]{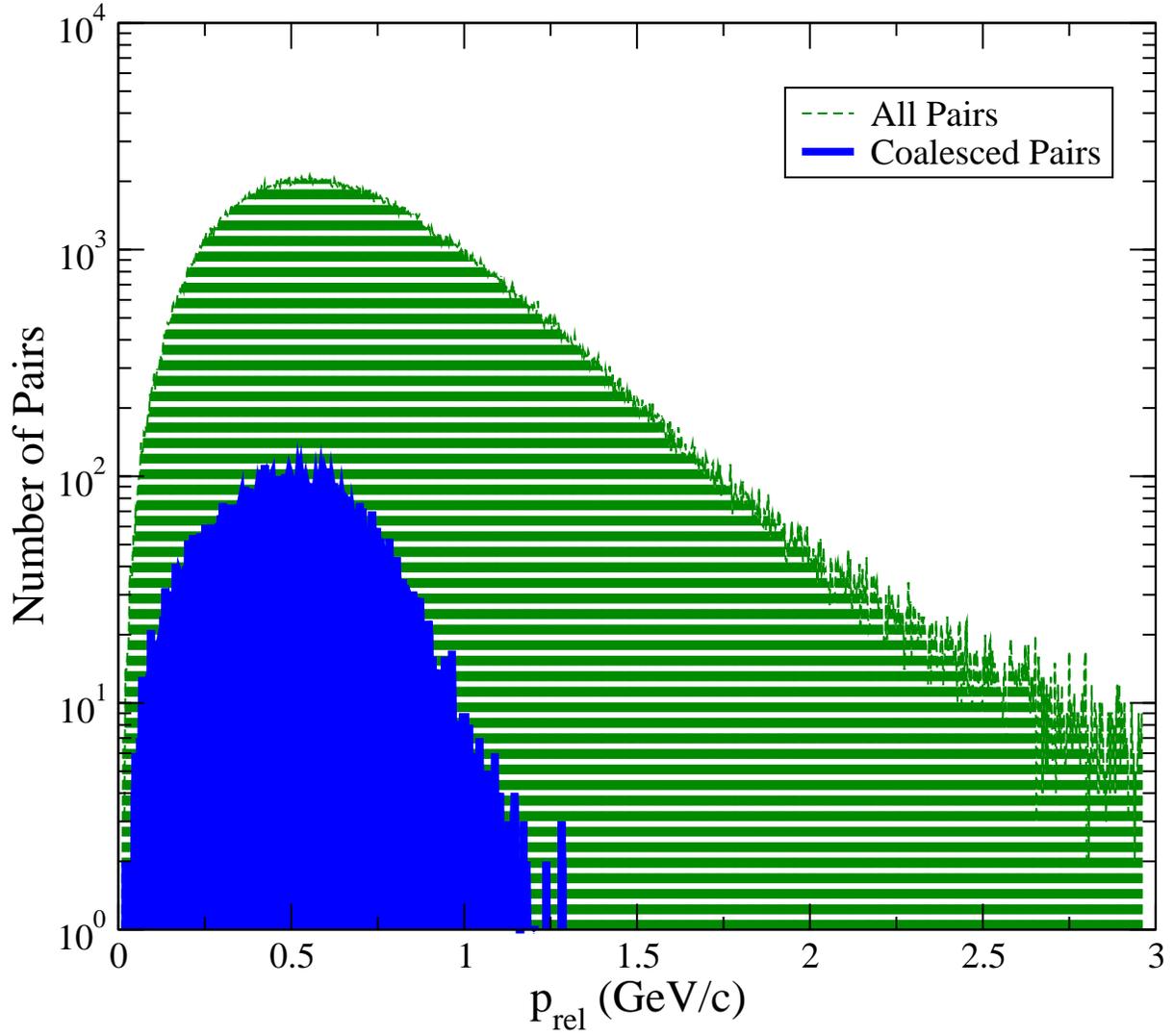}
\caption[]{Momentum distribution of coalesced $c\bar c$ pairs vs the same for
  all $c\bar c$ pairs in 200A GeV Au+Au collisions: a sharp cutoff in $c\bar
  c$ relative momentum is evident for the coalesced pairs, reflecting the
  structure of the charmonium wave functions taken from the Cornell model.}
\label{fig:Fig.(5)}
\end{figure}
\clearpage

\begin{figure}
%\vbox{\hbox to\hsize{\hfil
%\epsfxsize=6.1truein\epsffile[0 0 561 751]{figs/PUB.Journal.Direct.dndy.ps}
%\hfil}}
\includegraphics*[trim= -50 0 -10 -10,scale=1.0]{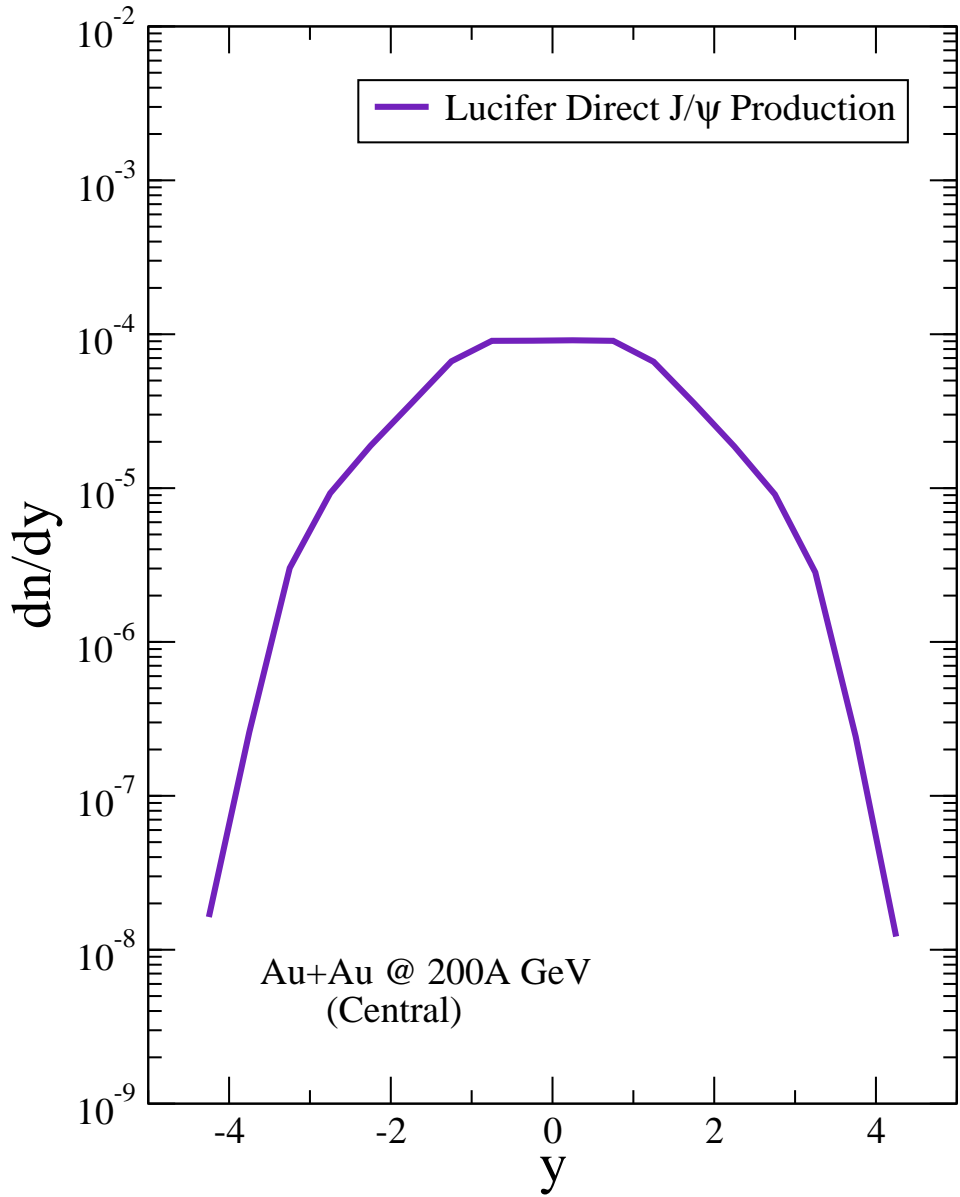}
\caption[]{Rapidity distribution for direct production of J/$\psi$.  Direct
  production involves the input of elementary hadron-hadron J/$\psi$
  cross-sections at a variety of energies.  As discussed in the text, the
  direct process is expected to be of less importance at sufficiently high
  collision energy. This is evidently already the case at the maximum RHIC
  energy of 200A GeV.}
\label{fig:Fig.(6)}
\end{figure}
\clearpage

\begin{figure}
%\vbox{\hbox to\hsize{\hfil
%\epsfxsize=6.1truein\epsffile[0 0 561 751]{figs/PUB.Journal.S0vsS1.jpsiINV.ps}
%\hfil}}
\includegraphics*[trim= -65 0 -10 -10,scale=1.0]{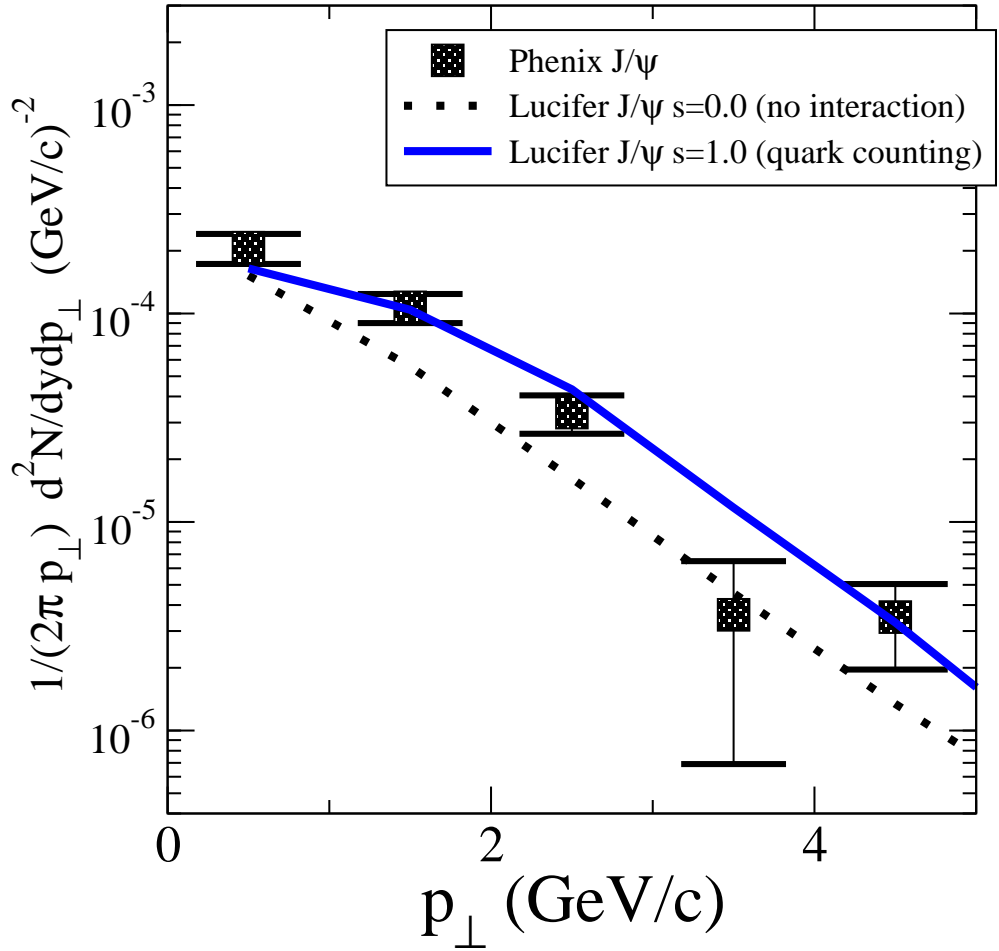}
\caption[]{Dependence of the J/$\psi$ tranverse momentum distribution on the
  strength of the in medium charm quark interactions. The quark-counting
  estimate ($s=1$) for the quark/pre-hadron interaction is clearly necessary
  to explain the relatively slow decreae with $p_\perp$ seen in the PHENIX
  data. For the purposes of this comparison, the $s=0$ curve has been
  normalized to agree with $s=1$ at low $p_\perp$, in order to highlight the
  differences in transverse momentum behaviour.  Charm quark interactions with
  the medium occur early on in our simulation, and this suggests that heavy
  quarkonia at the LHC may again provide a hadronic signal of the initial
  state in ion-ion collisions at such elevated energies.}
\label{fig:Fig.(7)}
\end{figure}

\end{document}